\documentclass{article}
\usepackage{spconf}
\usepackage{cite}
\usepackage{svg}
\usepackage{amsmath,amssymb,amsfonts}
\usepackage{algorithmic}
\usepackage{graphicx}
\usepackage{textcomp}
\usepackage{float}
\usepackage{xcolor}
\usepackage{stackengine}
\usepackage{multirow}
\usepackage{hhline}

\begin{document}\sloppy

\title{Texture-aware Video Frame Interpolation}
\name{Duolikun Danier and David Bull\thanks{This work was supported by the China Scholarship Council - University of Bristol Scholarship. Grant No. 202008060038.}}
\address{Visual Information Laboratory\\University of Bristol\\Bristol, BS8 1UB, United Kingdom \\ \{Duolikun.Danier, Dave.Bull\}@bristol.ac.uk}

\maketitle

\begin{abstract}
Temporal interpolation has the potential to be a powerful tool for video compression. Existing methods for frame interpolation do not discriminate between video textures and generally invoke a single general model capable of interpolating a wide range of video content. However, past work on video texture analysis and synthesis has shown that different textures exhibit vastly different motion characteristics and they can be divided into three classes (static, dynamic continuous and dynamic discrete). In this work, we study the impact of video textures on video frame interpolation, and propose a novel framework where, given an interpolation algorithm, separate models are trained on different textures. Our study shows that video texture has significant impact on the performance of frame interpolation models and it is beneficial to have separate models specifically adapted to these  texture classes, instead of training a single model that tries to learn generic motion. Our results demonstrate that models fine-tuned using our framework achieve, on average, a 0.3dB gain in PSNR on the test set used.
\end{abstract}


\section{Introduction}
Video frame interpolation (VFI) refers to the task of generating non-existent intermediate frames between any two consecutive frames in a video while preserving spatiotemporal consistencies. VFI allows up-conversion of frame rates for improved visual quality, and offers an important tool for video compression, where it can be used to replace or enhance conventional motion estimation\cite{b1} or to conceal errors in reconstructed frames\cite{b2}.
Interpolating visually pleasing intermediate frames requires accurate modelling of motion. However, due to the variety of textures and motion patterns in real-world videos, it is difficult to capture the motion of different texture types using a single mathematical model. Hence frame interpolation remains a challenging task. 

Existing VFI methods can be classified as either flow-based or kernel-based. Flow-based methods\cite{superslomo, baker2011, dvf, toflow, ctx, dain, softsplat} generally involve two steps, namely motion estimation and frame synthesis, where the latter typically resorts to pixel-wise warping of the two adjacent frames utilising the motion information to predict the intermediate frame. However, these methods generally experience a degradation in performance under challenging conditions where finding reference pixels in nearby frames is made difficult, e.g. under illumination change, occlusion and large motion. On the other hand, kernel-based methods\cite{long, adaconv, sepconv, cain, revisit, adacof, dsepconv, gdconv} estimate a convolution kernel for each output pixel and predict the pixel by convolving nearby regions in adjacent frames with the kernel. Although in this case each output pixel is synthesised by referring to multiple pixels in the adjacent frames and occlusion can be better handled, the complexity of captured motion is still limited by the size of the kernels. 

The aforementioned methods are mostly based on convolutional neural networks (CNNs). These have advanced the state-of-the-art in VFI over the past few years with various innovations relating to their overall architectures and to specific processing stages. One common feature amongst them is that a single model is trained to interpolate all kinds of video frame. In contrast, previous work on video texture analysis and synthesis\cite{parametric, dyntex, afonso, understanding, bvitexture, syntex} has shown that different textures exhibit vastly different motion patterns which are better handled separately. For example, Zhang et al.\cite{parametric} proposed an analysis-synthesis video compression framework in which different algorithms were adopted to synthesise dynamic and static textures.

\begin{figure}[t]
\centering 
\includegraphics[width=0.15\textwidth]{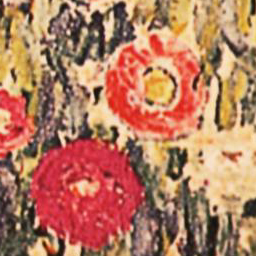}
\includegraphics[width=0.15\textwidth]{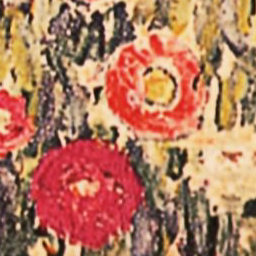}
\includegraphics[width=0.15\textwidth]{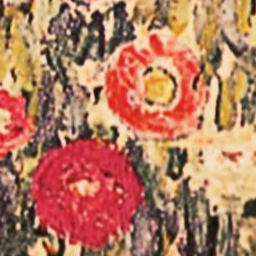}

\includegraphics[width=0.15\textwidth]{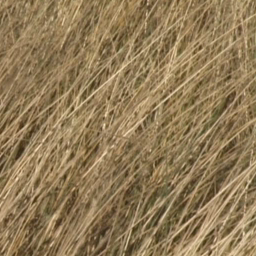}
\includegraphics[width=0.15\textwidth]{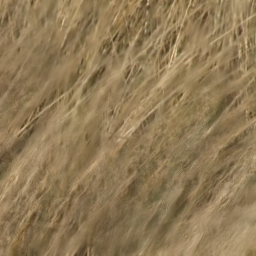}
\includegraphics[width=0.15\textwidth]{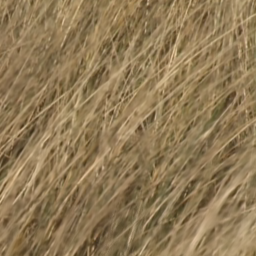}

\stackunder[2pt]{\includegraphics[width=0.15\textwidth]{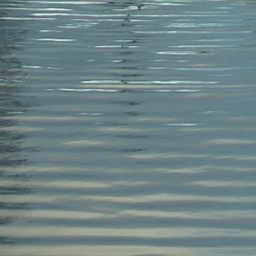}}{\begin{small}Original frames\end{small}}
\stackunder[2pt]{\includegraphics[width=0.15\textwidth]{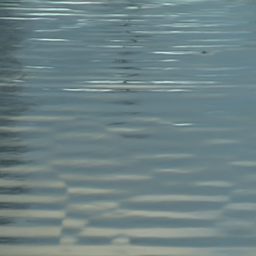}}{\begin{small}AdaCoF-original\end{small}}
\stackunder[2pt]{\includegraphics[width=0.15\textwidth]{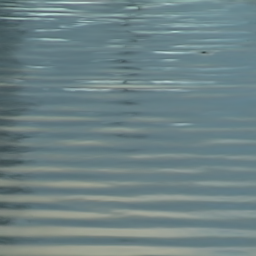}}{\begin{small}AdaCoF-TAFI\end{small}}
\caption{Interpolation results of our texture-aware fine-tuned AdaCoF and the original AdaCoF\cite{adacof} on sample sequences from static (top row), dynamic discrete (middle row) and dynamic continuous (bottom row) textures. Sequences are ``PaintingTilting1", ``RiceField" and ``ShinnyBlueWater\_downsampled" from the HomTex dataset\cite{afonso}.}
\label{example}
\end{figure}

A finer classification of video textures was proposed in \cite{afonso, understanding}, where the authors clustered homogeneous sequences based on texture-relevant features as well as the HM encoding statistics and found three video texture clusters exist which were referred to as \textit{static} (rigid texture exhibiting perspective motion), \textit{dynamic discrete} (texture with discernible parts undergoing perspective motion independently) and \textit{dynamic continuous} (spatially irregular and unstructured texture moving as a continuum). Example frames from each class taken from the HomTex dataset\cite{afonso} are shown in Fig.\ref{example}.

While generalisation is a desirable property of deep learning models, it has also proved to be difficult due to the issue of over-fitting. Motivated by the classification of video textures in previous work, in this contribution, we investigate the impact of video texture type on the performance of state-of-the-art VFI models. Based on our observation, we hypothesise that it is harder to train a generic VFI model that performs well on all texture types than to train separate models, each specialising in a specific texture (i.e. to overfit texture class); we verify our hypothesis through experimentation. As a consequence,  we propose a novel texture-aware frame interpolation (TAFI) framework that can generalise to any VFI method and improve its performance. 

\section{Related Work}
Conventionally, VFI has been addressed by estimating optical flow and warping two adjacent frames using the flow map\cite{baker2011}. Such methods rely heavily on the adopted flow estimation algorithm and have been generally found to suffer from occlusion and large motion. Recent flow-based methods improved upon the framework by deploying CNNs for flow estimation and frame synthesis. Liu et al.\cite{dvf} trained an encoder-decoder network to estimate the optical flow from the intermediate frame to the two reference frames, which is then used by a sampling layer to interpolate each output pixel. The complexity of the captured motion was limited since equal flow in two directions was assumed. Jiang et al.\cite{superslomo} improved on \cite{dvf} by enabling bidirectional flow. Niklaus et al.\cite{ctx} incorporated context information during the warping process. Bao et al.\cite{dain} additionally estimated depth map to refine the estimated flow. Recently, Niklaus et al.\cite{softsplat} proposed to use softmax splatting for differentiable forward warping such that different reference pixels can map to the same output pixel. Although this allows more complex motion to be captured, such complexity is still limited due to the dependence on flow estimation and pixel-wise warping. 

On the other hand, Kernel-based methods inherently allow more pixels to be sampled for predicting one output pixel. Niklaus et al.\cite{adaconv} proposed a CNN (improved later in \cite{sepconv} and \cite{revisit}) to predict a kernel for each output pixel which is convolved with the input image to obtain the target pixel. Instead of predicting individual kernels for each pixel, Choi et al.\cite{cain} proposed an end-to-end network with channel attention that directly outputs the interpolated frame. Recently, there has been an increased number of works on VFI that employs deformable convolution\cite{defconv}. Lee et al.\cite{adacof} developed a CNN to predict the kernel weights together with their offsets for each output pixel. The flexibility of deformable kernels enabled more complex motion to be captured. Cheng et al.\cite{dsepconv} adopted a similar approach but separable kernels were used. Shi et al.\cite{gdconv} enabled further degree of freedom by allowing the reference pixels to be sampled in an interpolated space-time volume instead of just from the existent frames.

While these VFI methods were developed to generalise on all types of videos, previous work on video texture analysis and synthesis adopted a different approach. In\cite{parametric}, Zhang et al. observed that textured regions in videos typically consume more bits to encode, and proposed to synthesise such region and encode only the warping parameters, where different synthesis methods were used for dynamic and static textures. In \cite{afonso}, Afonso et al. analysed video texture by clustering homogeneous videos based on their encoding statistics (e.g. prediction modes, bit allocation etc.) and showed that video texture can be categorised as \textit{static}, \textit{dynamic discrete} and \textit{dynamic continuous}. A homogeneous texture dataset, HomTex, was produced. Katsenou et al.\cite{understanding} confirmed the existence of such categorisation by clustering videos based on their spatiotemporal features. In our work, the implications of such texture categorisation for VFI are investigated.

\section{Texture Aware Frame Interpolation}
\begin{figure*}[t]
\centerline{\includegraphics[width=\linewidth]{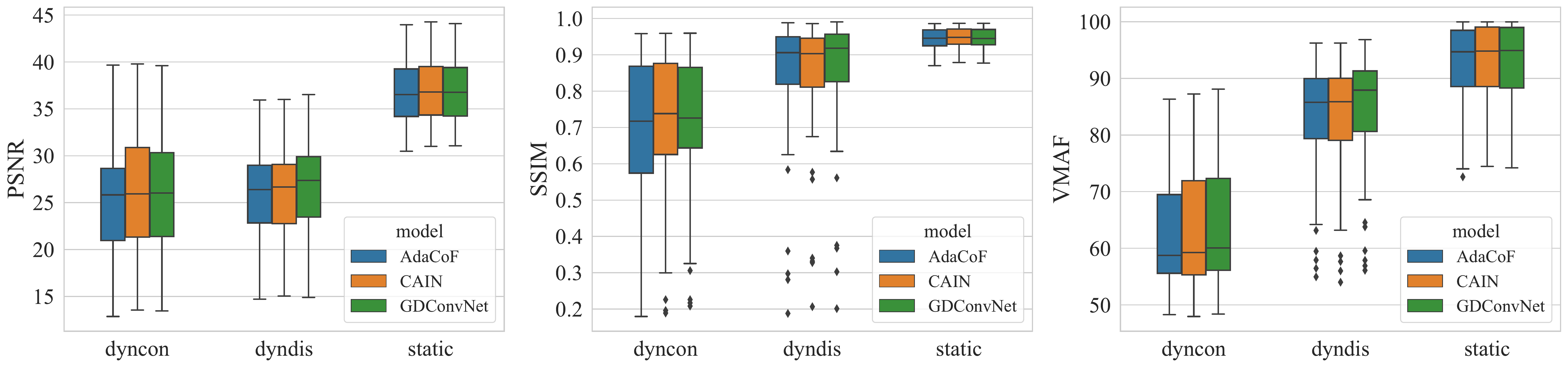}}
\vspace{-1em}
\caption{Performance of three selected models on three texture types.}
\label{boxplot}
\end{figure*}
\subsection{Effect of Texture on VFI Performance}
In this section we investigate the performance of VFI methods on different video textures. Specifically, the pre-trained versions of three CNN-based state-of-the-art VFI models with publicly available source code are evaluated: AdaCoF\cite{adacof}, CAIN\cite{cain} and GDConvNet\cite{gdconv}. The evaluation dataset used is HomTex\cite{afonso} which contains 120 videos with 250 frames of $256\times 256$ spatial resolution at 25 and 60 fps. Each video in HomTex is (approximately) texturally homogeneous;  there are 45, 50 and 25 sequences that are labelled \textit{dynamic continuous}, \textit{dynamic discrete} and \textit{static} respectively. Each model is evaluated on the entire HomTex dataset, where for each sequence every second frame, $I_t$, is considered the ground-truth and the two adjacent frames $I_{t-1}$ and $I_{t+1}$ are input to the model (in the case of GDConvNet 4 frames, namely $I_{t-3}, I_{t-1}, I_{t+1}, I_{t+3}$, are used as input due to the model design). Peak signal-to-noise ratio (PSNR) and structural similarity (SSIM)\cite{ssim} scores are computed using the original and interpolated frames as these are the most commonly used metrics in VFI\cite{superslomo, baker2011, dvf, toflow, ctx, dain, softsplat, adacof, long, adaconv, sepconv, cain, revisit, dsepconv, gdconv}. Additionally, video multi-method assessment fusion (VMAF)\cite{vmaf} scores are also computed using the original and interpolated videos. The scores of each model are then grouped according to the video texture class, and their distributions are shown in Fig.\ref{boxplot}, where abbreviations ``dyncon" and ``dyndis" are used for \textit{dynamic continuous} and \textit{dynamic discrete}. 

It is noticeable that the distributions of the scores obtained by the models vary with texture type. Specifically, all models scored the highest PSNR, SSIM and VMAF on static textures, which is expected since the perspective motion of rigid objects is relatively easier to capture. For the case of dynamic texture, all models see degraded performance across all metrics, and this decrease is even more severe for dynamic continuous textures which exhibit the most irregular motion intrinsic to the scenes containing water, smoke and fire, etc.
\begin{table}[t]
\begin{center}
\label{anova}
\caption{Results of One-way ANOVA test on PSNR, SSIM and VMAF scores obtained by the VFI models on different textures.}
\vspace{-1em}
\resizebox{\linewidth}{!}{\begin{tabular}{l|cc|cc|cc}
\hline
     & \multicolumn{2}{c|}{AdaCoF} & \multicolumn{2}{c|}{CAIN} & \multicolumn{2}{c}{GDConvNet} \\ \hline
     & F(2,117)       & p          & F(2,117)      & p         & F(2,117)        & p           \\ \hline
PSNR & 38.41          & 0.00       & 40.14         & 0.00      & 35.99           & 0.00        \\
SSIM & 17.58          & 0.00       & 16.16         & 0.00      & 16.91           & 0.00        \\
VMAF & 73.57          & 0.00       & 67.92         & 0.00      & 70.07           & 0.00       \\ \hline
\end{tabular}}
\end{center}
\end{table}

To assess whether the impact of texture on VFI performance is statistically significant, a one-way ANOVA is performed with results shown in Table~\ref{anova}. Here we see that in terms of all metrics, texture type has significant effect on the performance of the VFI models at $p<0.05$ level. Furthermore, T-tests (Welch's T-test to account for unequal sample sizes) are performed between the scores of each model on each pair of textures to see if there are significant differences, and the results are summarised in Table~\ref{ttest}. The last two rows of the table show that, in terms of all metrics, the average performance of the models on static textures is significantly different from that on dynamic discrete and dynamic continuous textures. Comparing between the two dynamic types, all models have significantly different SSIM and VMAF scores. Combining these results with Fig.\ref{boxplot}, we can confirm that for $p<0.05$, all models performed best on static textures and worst on dynamic continuous textures. There are two possible reasons for this observation. The first is that the dynamic texture types fall outside of the distribution of the three models' training sets. We consider this reason less important since all models were trained on the Vimeo90K dataset\cite{toflow} which involves a wide range of non-homogeneous videos that cover all types of textures. The second reason is that the complexity of motion patterns, and hence the levels of difficulties to interpolate different textures are inherently different, with dynamic continuous textures exhibiting the most unpredictable motion thus being the hardest to interpolate, and static textures being the easiest type. This implies that, rather than trying to train a single model that learns to model the motion of all textures, there might be potential gain in having separate models specialised in interpolation of one texture class, namely texture-aware frame interpolation. 

\begin{table*}[t]
\caption{Results of T-Tests between the scores obtained by each model on three pairs of textures.}
\vspace{-1em}
\begin{center}
\resizebox{\textwidth}{!}{\begin{tabular}{c|ccc|ccc|ccc}
\hline
\multicolumn{1}{l|}{}                                        & \multicolumn{3}{c|}{AdaCoF}                                                                                                                                                                  & \multicolumn{3}{c|}{CAIN}                                                                                                                                                                     & \multicolumn{3}{c}{GDConvNet}                                                                                                                                                                \\ \cline{2-10} 
\multicolumn{1}{l|}{}                                        & PSNR                                                          & SSIM                                                         & VMAF                                                          & PSNR                                                          & SSIM                                                         & VMAF                                                           & PSNR                                                          & SSIM                                                         & VMAF                                                          \\ \hline
\begin{tabular}[c]{@{}c@{}}dyndis vs. \\ dyncon\end{tabular} & \begin{tabular}[c]{@{}c@{}}t(75)=0.10,\\ p=0.92\end{tabular}  & \begin{tabular}[c]{@{}c@{}}t(87)=3.52,\\ p=0.00\end{tabular} & \begin{tabular}[c]{@{}c@{}}t(92)=8.93,\\ p=0.00\end{tabular}  & \begin{tabular}[c]{@{}c@{}}t(73)=-0.19,\\ p=0.85\end{tabular} & \begin{tabular}[c]{@{}c@{}}t(87)=3.21,\\ p=0.00\end{tabular} & \begin{tabular}[c]{@{}c@{}}t(92)=8.35,\\ p=0.00\end{tabular}   & \begin{tabular}[c]{@{}c@{}}t(74)=0.18,\\ p=0.86\end{tabular}  & \begin{tabular}[c]{@{}c@{}}t(87)=3.48,\\ p=0.00\end{tabular} & \begin{tabular}[c]{@{}c@{}}t(91)=8.90,\\ p=0.00\end{tabular}  \\ \hline
\begin{tabular}[c]{@{}c@{}}static vs. \\ dyncon\end{tabular} & \begin{tabular}[c]{@{}c@{}}t(66)=8.66,\\ p=0.00\end{tabular}  & \begin{tabular}[c]{@{}c@{}}t(47)=7.78,\\ p=0.00\end{tabular} & \begin{tabular}[c]{@{}c@{}}t(58)=12.28,\\ p=0.00\end{tabular} & \begin{tabular}[c]{@{}c@{}}t(66)=8.64,\\ p=0.00\end{tabular}  & \begin{tabular}[c]{@{}c@{}}t(46)=7.53,\\ p=0.00\end{tabular} & \begin{tabular}[c]{@{}c@{}}t(61)=12.08, \\ p=0.00\end{tabular} & \begin{tabular}[c]{@{}c@{}}t(66)=8.47,\\ p=0.00\end{tabular}  & \begin{tabular}[c]{@{}c@{}}t(46)=7.63,\\ p=0.00\end{tabular} & \begin{tabular}[c]{@{}c@{}}t(61)=11.85,\\ p=0.00\end{tabular} \\ \hline
\begin{tabular}[c]{@{}c@{}}static vs. \\ dyndis\end{tabular} & \begin{tabular}[c]{@{}c@{}}t(62)=11.41,\\ p=0.00\end{tabular} & \begin{tabular}[c]{@{}c@{}}t(54)=4.01,\\ p=0.00\end{tabular} & \begin{tabular}[c]{@{}c@{}}t(59)=3.84,\\ p=0.00\end{tabular}  & \begin{tabular}[c]{@{}c@{}}t(61)=11.91,\\ p=0.00\end{tabular} & \begin{tabular}[c]{@{}c@{}}t(53)=4.24,\\ p=0.00\end{tabular} & \begin{tabular}[c]{@{}c@{}}t(61)=4.06,\\ p=0.00\end{tabular}   & \begin{tabular}[c]{@{}c@{}}t(63)=11.03,\\ p=0.00\end{tabular} & \begin{tabular}[c]{@{}c@{}}t(53)=3.89,\\ p=0.00\end{tabular} & \begin{tabular}[c]{@{}c@{}}t(60)=3.40,\\ p=0.00\end{tabular}  \\ \hline
\end{tabular}}
\label{ttest}
\end{center}
\end{table*}

\subsection{Proposed Method}
Under the proposed VFI framework, the aforementioned three texture types are treated separately. That is, given a VFI model, we generate three versions of it, each trained exclusively on one type of texture. In this work we focus on performance of models on homogeneously-textured videos so the inference on a test video is performed in the same way as the original model, but with the difference that the specialised versions are used according to the video texture type. We show through our experiments that it is more difficult to learn a generic model that can interpolate all types of videos than to learn a specialised model that overfits a specific texture class. 

\section{Experiments}
\subsection{Experiment Setup}
\subsubsection{Datasets}
We focus on the three models used previously for analysis, i.e. AdaCoF\cite{adacof}, CAIN\cite{cain} and GDConvNet\cite{gdconv}. The test data used for evaluation are the 120 homogeneous sequences in HomTex\cite{afonso}. For the purpose of training models tuned to a single texture type, the training dataset should contain homogeneous sequences. Therefore, we make use of three existing texture datasets, namely DynTex\cite{dyntex}, BVI-Texture\cite{bvitexture} and SynTex\cite{syntex}. Specifically, DynTex contains 650 annotated videos of $720\times 576$ spatial resolution at 25 fps, and we retained those that contain only one type of texture according to the annotations. BVI-Texture and SynTex each contain 20 and 196 full-HD resolution ($1920\times 1080)$ homogeneous sequences at 60 fps. It should be noted that, since HomTex is composed of videos from DynTex and BVI-Texture, we removed the DynTex sequences that exist in HomTex and other sequences that are visually similar to them. This resulted in a total of 214 dynamic continuous, 222 dynamic discrete and 110 static sequences with each sequence containing at least 250 frames.

\subsubsection{Training details} Due to the lack of a diverse homogeneous video texture dataset, the models are initialised with their pre-trained weights provided and fine-tuned on our training sets. The loss functions, optimisation strategies and other hyperparameters except learning rate are set identical to the original implementations of the models. To fine-tune a model on a specific texture class, each training batch is formed by randomly sampling 10000 triplets from videos of that texture where each time a sequence is randomly selected then a triplet (or quintuplet for GDConvNet) of $256\times 256$ patches is sampled randomly from the space-time volume. The sampled frames are augmented via random horizontal and vertical flipping, color jittering and temporal order reversal. Batch sizes of $8,8,3$ are used and the initial learning rates are set to be $10^{-3}, 10^{-5}, 10^{-4}$ for AdaCoF, CAIN and GDConvNet respectively. The models are fine-tuned on each video texture for 10 epochs with learning rates decayed by a factor of $0.5$ every 4 epochs. The computations are performed on Nvidia P100 GPU cards provided on the shared cluster BlueCrystal Phase 4\cite{bc4} at the University of Bristol.

\subsection{Effectiveness of TAFI}
\begin{table*}[t]
\setlength\abovecaptionskip{-0.7\baselineskip}
\caption{Performance of the original (baseline) and texture-aware fine-tuned versions of AdaCoF, CAIN and GDConvNet on HomTex. The version indicates the texture the model is fine-tuned on. The combination of dyncon, dyndis and static versions forms the TAFI model. Models are evaluated on three individual texture types in HomTex as well as the whole HomTex dataset. Numbers in brackets denote change with respect to the baseline model. For each column, the best result is in bold text. Our TAFI framework consistently improved the overall performance of the baseline models.}
\vspace{-1.5em}
\begin{center}
\resizebox{\textwidth}{!}{\begin{tabular}{cl|ll|ll|ll|ll}
\hline
 & \multicolumn{1}{c|}{} & \multicolumn{2}{c|}{HomTex-dyncon} & \multicolumn{2}{c|}{HomTex-dyndis} & \multicolumn{2}{c|}{HomTex-static} & \multicolumn{2}{c}{HomTex-overall} \\ \hline 
\multicolumn{1}{c|}{model} & \multicolumn{1}{c|}{version} & \multicolumn{1}{c}{PSNR} & \multicolumn{1}{c|}{SSIM} & \multicolumn{1}{c}{PSNR} & \multicolumn{1}{c|}{SSIM} & \multicolumn{1}{c}{PSNR} & \multicolumn{1}{c|}{SSIM} & \multicolumn{1}{c}{PSNR} & \multicolumn{1}{c}{SSIM}  \\ \hline
\multicolumn{1}{c|}{\multirow{5}{*}{AdaCoF}} & baseline & 25.90 & 0.68 & 26.03 & 0.83 & {36.69} & {0.94} & 28.20 & 0.80 \\ \cline{2-10} 
\multicolumn{1}{c|}{} & dyncon (ours) & \textbf{26.36(+0.46)} & \textbf{0.70(+0.02)} & 25.87(-0.16) & 0.83(+0.00) & 34.72(-1.97) & 0.89(-0.05) & \multirow{3}{*}{\textbf{28.51(+0.31)}} & \multirow{3}{*}{\textbf{0.81(+0.01)}}\\
\multicolumn{1}{c|}{} & dyndis (ours) & 26.14(+0.24) & 0.69(+0.01) & \textbf{26.16(+0.13)} & \textbf{0.84(+0.01)} & 35.40(-1.29) & 0.92(-0.02) & &\\
\multicolumn{1}{c|}{} & static (ours) & 25.79(-0.11) & 0.67(-0.01) & 25.99(-0.04) & 0.83(+0.00) & \textbf{37.11(+0.42)} & \textbf{0.94(+0.00)} & & \\ \cline{2-10} 
\multicolumn{1}{c|}{} & mixed & {26.19(+0.29)} & {0.69(+0.01)} & {26.12(+0.09)} & {0.84(+0.01)} & 36.01(-0.68) & 0.93(-0.01) & 28.22(+0.02) & 0.80(+0.00) \\ \hline \hline
\multicolumn{1}{c|}{\multirow{5}{*}{CAIN}} & baseline & 26.26 & 0.70 & {26.03} & 0.83 & {36.96} & {0.95} & 28.39 & 0.81 \\ \cline{2-10}
\multicolumn{1}{c|}{} & dyncon (ours) & \textbf{26.63(+0.37)} & \textbf{0.70(+0.00)} & 25.58(-0.45) & 0.82(-0.01) & 35.21(-1.75) & 0.92(-0.03) & \multirow{3}{*}{\textbf{28.69(+0.30)}} & \multirow{3}{*}{\textbf{0.81(+0.00)}} \\
\multicolumn{1}{c|}{} & dyndis (ours) & {26.44(+0.18)} & 0.70(-0.00) & \textbf{26.33(+0.30)} & \textbf{0.84(+0.01)} & 35.90(-1.06) & 0.93(-0.02) & & \\
\multicolumn{1}{c|}{} & static (ours) & 26.17(-0.09) & 0.69(-0.01) & 25.92(-0.11) & 0.83(-0.00) & \textbf{37.14(+0.18)} & \textbf{0.96(+0.01)} & & \\ \cline{2-10}
\multicolumn{1}{c|}{} & mixed & 26.42(+0.16) & {0.70(+0.00)} & 25.84(-0.19) & {0.84(+0.01)} & 36.07(-0.89) & 0.94(-0.01) & 28.19(-0.20) & 0.81(+0.00) \\ \hline \hline
\multicolumn{1}{c|}{\multirow{5}{*}{GDConvNet}} & baseline &  26.38 & 0.70 & 26.60 & 0.84  & {36.97} & {0.95} & 28.68 & 0.81 \\ \cline{2-10}
\multicolumn{1}{c|}{} & dyncon (ours) & \textbf{26.76(+0.38)} & \textbf{0.72(+0.02)} & 25.78(-0.82) & 0.82(-0.02) & 33.86(-3.11) & 0.88(-0.07) & \multirow{3}{*}{\textbf{28.97(+0.29)}} & \multirow{3}{*}{\textbf{0.82(+0.01)}} \\ 
\multicolumn{1}{c|}{} & dyndis (ours) & 26.02(-0.36) & 0.69(-0.01) & \textbf{26.89(+0.29)} & \textbf{0.85(+0.01)} & 33.58(-3.39) & 0.87(-0.08) & & \\
\multicolumn{1}{c|}{} & static (ours) & 25.56(-0.82) & 0.66(-0.04) & 25.43(-1.17) & 0.82(-0.02) & \textbf{37.10(+0.13)} & \textbf{0.96(+0.01)} & & \\ \cline{2-10}
\multicolumn{1}{c|}{} & mixed & {26.48(+0.10)} & {0.70(+0.00)} &{26.72(+0.12)} & {0.84(+0.00)} & 35.38(-1.59) & 0.93(-0.02) & 28.43(-0.25) & 0.81(-0.00) \\ \hline
\end{tabular}}
\label{final}
\end{center}
\end{table*}
In order to investigate whether there is benefit in texture-specific training instead of training the models on all texture classes, in this experiment we fine-tune each model on four training sets: three for the three texture classes as described above and also the combination of them (referred to as ``mixed"). We name the fine-tuned versions after the targetted texture type, namely ``dyncon", ``dyndis", ``static" and ``mixed". The four fine-tuned versions, as well as the original ``off-the-shelf" version (regarded as the baseline) of each model are then evaluated on HomTex. The average PSNR and SSIM scores obtained for each texture subset in HomTex (i.e. HomTex-dyncon, HomTex-dyndis and HomTex-static) and the overall HomTex dataset (HomTex-overall) are summarised in Table~\ref{final}.

We see from Table~\ref{final} that for all baseline models, texture class tuning increased their performance on test sequences from that class in terms of both PSNR and SSIM, although in general the variations in SSIM scores are marginal. Meanwhile, models fine-tuned on a class generally performed worse on other textures, implying some overfitting which is expected. The improvements due to class-based training are particularly evident for dynamic textures. 

The performance variations are more obviously reflected by PSNR, and this is because all models use $\ell 1$ distortion as a major component in their loss functions, so one can expect a lower mean-squared-error after training and hence the obvious variations in PSNR. 

Comparing the specialised models that are fine-tuned on single texture types (which together form the TAFI model) against the model fine-tuned on the mixed training set, we see that in all cases the former managed to deliver increased gain in terms of both PSNR and SSIM. This result is consistent with our hypothesis that learning a generic model is relatively harder than specialising on a specific video texture class. It is noted that the models fine-tuned on the mixed set constantly exhibited worse performance on static textures while performance on the other two texture types are mostly improved. This can be attributed to the fact that the static sequences in the training set constitute a relatively small proportion (20\%) in the mixed set, making the training more biased towards the dynamic texture types.

Finally, comparing the performance of the models on the entire HomTex dataset, it is clear that for all three VFI models, the combination of their specialised versions (TAFI) achieved higher PSNR and SSIM than their baselines, with approximately 0.3dB gain in PSNR after texture-aware fine-tuning. Example qualitative interpolation results of the specialised AdaCoF models and the original AdaCoF are given in Fig.\ref{example}, where it can be observed that the specialised versions (AdaCoF-TAFI) produce an intermediate frame with higher visual quality. This is particularly evident for dynamic textures where the original AdaCoF model fails to capture the more complex motion and produces numerous visual artefacts.

\section{Conclusion and Future Work}
In this work, we applied the video texture categorisation proposed in previous work to the problem of video frame interpolation (VFI). The effect of video texture type on VFI was studied by evaluating three state-of-the-art models on a video dataset HomTex which contains homogeneous sequences of static, dynamic discrete and dynamic continuous textures. The results showed that the models perform differently on different textures with statistical significance. Motivated by this observation, we fine-tuned the three VFI models on each of the three types of textures as well as the combination of them, and found that although each model fine-tuned on a specific texture overfits that texture class, it outperformed the model trained on combination of textures, proving our hypothesis that it is beneficial in terms of interpolation quality to have separate models specialised in different textures, instead of training a single model to learn to interpolate all kinds of textures. 

To further confirm these findings, VFI models should be completely re-trained on larger scale homogeneous video datasets which do not yet exist. A possible direction of future work is to construct such a dataset and train state-of-the-art VFI models on homogeneous videos from scratch. In addition, despite the improved performance, our framework requires three times more computation compared to the original VFI models and there is up to future research to develop a single model that can adapt to different textures. Finally, in this work we focused on performance of VFI on homogeneous videos and did not consider fusion of the specialised models to interpolate generic sequences. Development of such ensemble algorithms will also be part of our future work.


\begin{thebibliography}{00}
\bibitem{b1} H. Choi and I. V. Bajić, ``Deep frame prediction for video coding," \textit{IEEE Transactions on Circuits and Systems for Video Technology}, vol. 30, no. 7, pp. 1843-1855, July 2020.
\bibitem{b2} M. Usman, X. He, K. Lam, M. Xu, S. M. M. Bokhari and J. Chen, ``Frame Interpolation for Cloud-Based Mobile Video Streaming," \textit{IEEE Transactions on Multimedia}, vol. 18, no. 5, pp. 831-839, May 2016.
\bibitem{baker2011} S. Baker, S. Roth, D. Scharstein, M. J. Black, J. P. Lewis and R. Szeliski, ``A database and evaluation methodology for optical flow," in \textit{IEEE 11th International Conference on Computer Vision}, 2007, pp. 1-8.
\bibitem{dvf} Z. Liu, R. A. Yeh, X. Tang, Y. Liu and A. Agarwala, ``Video frame synthesis using deep voxel flow," in \textit{IEEE International Conference on Computer Vision}, 2017, pp. 4473-4481.
\bibitem{superslomo} H. Jiang, D. Sun, V. Jampani, M. Yang, E. Learned-Miller and J. Kautz,``Super SloMo: high quality estimation of multiple intermediate frames for video interpolation," in \textit{IEEE Conference on Computer Vision and Pattern Recognition}, 2018, pp. 9000-9008.
\bibitem{ctx} S. Niklaus and F. Liu, ``Context-aware synthesis for video frame interpolation," in \textit{IEEE Conference on Computer Vision and Pattern Recognition}, 2018, pp. 1701-1710.
\bibitem{dain} W. Bao, W. Lai, C. Ma, X. Zhang, Z. Gao and M. Yang, ``Depth-aware video frame interpolation," in \textit{IEEE Conference on Computer Vision and Pattern Recognition}, 2019, pp. 3698-3707.
\bibitem{toflow} T. Xue, B. Chen, J. Wu, D. Wei and W. T. Freeman, ``Video enhancement with task-oriented flow," \textit{International Journal of Computer Vision}, vol. 127, no. 8, pp. 1106-1125, 2019.
\bibitem{softsplat} S. Niklaus and F. Liu, ``Softmax splatting for video frame interpolation," in \textit{IEEE Conference on Computer Vision and Pattern Recognition}, 2020, pp. 5436-5445.
\bibitem{long} G. Long, L. Kneip, J. M. Alvarez, H. Li, X. Zhang, and Q. Yu, ``Learning image matching by simply watching video," in \textit{European Conference on Computer Vision}, 2016, pp. 434-450. Springer, Cham.
\bibitem{adaconv} S. Niklaus, L. Mai and F. Liu, ``Video frame interpolation via adaptive convolution," in \textit{IEEE Conference on Computer Vision and Pattern Recognition}, 2017, pp. 2270-2279.
\bibitem{sepconv} S. Niklaus, L. Mai and F. Liu, ``Video frame interpolation via adaptive separable convolution," in \textit{IEEE International Conference on Computer Vision}, 2017, pp. 261-270.
\bibitem{revisit} S. Niklaus, L. Mai and O. Wang, ``Revisiting adaptive convolutions for video frame interpolation," in \textit{Proceedings of the IEEE/CVF Winter Conference on Applications of Computer Vision}, 2021, pp. 1099-1109.
\bibitem{cain} M. Choi, H. Kim, B. Han, N. Xu, and K. M. Lee, ``Channel attention is all you need for video frame interpolation," in \textit{Proceedings of the AAAI Conference on Artificial Intelligence}, vol. 34, no. 7, pp. 10663-10671, April 2020.
\bibitem{adacof} H. Lee, T. Kim, T. -y. Chung, D. Pak, Y. Ban and S. Lee, ``AdaCoF: adaptive collaboration of flows for video frame interpolation," in \textit{IEEE Conference on Computer Vision and Pattern Recognition}, 2020, pp. 5315-5324.
\bibitem{dsepconv} X. Cheng and Z. Chen, ``Video frame interpolation via deformable separable convolution," in \textit{Proceedings of the AAAI Conference on Artificial Intelligence}, vol. 34, no. 07, pp. 10607-10614, April 2020.
\bibitem{gdconv} Z. Shi, X. Liu, K. Shi, L. Dai and J. Chen, ``Video interpolation via generalized deformable convolution," 2020, arXiv:2008.10680. [Online]. Available: https://arxiv.org/abs/2008.10680.
\bibitem{defconv} J. Dai, H. Qi, Y. Xiong, Y. Li, G. Zhang, H. Hu, and Y. Wei, ``Deformable convolutional networks," in \textit{IEEE International Conference on Computer Vision}, 2017, pp. 764-773.
\bibitem{parametric} F. Zhang and  D. R. Bull, ``A parametric framework for video compression using region-based texture models," \textit{IEEE Journal of Selected Topics in Signal Processing}, vol. 5, no. 7, pp. 1378-1392, 2011.
\bibitem{dyntex} R. P\'eteri, S. Fazekas, and M. J. Huiskes, ``Dyntex: A comprehensive database of dynamic textures,” \textit{Pattern Recognition Letters}, vol. 31, no. 12, pp. 1627–1632, 2010.
\bibitem{afonso} M. Afonso, A. Katsenou, F. Zhang, D. Agrafiotis and D. Bull, ``Video texture analysis based on HEVC encoding statistics," in \textit{Picture Coding Symposium}, 2016, pp. 1-5.
\bibitem{understanding} A. V. Katsenou, T. Ntasios, M. Afonso, D. Agrafiotis and D. R. Bull, ``Understanding video texture — A basis for video compression," \textit{IEEE 19th International Workshop on Multimedia Signal Processing}, 2017, pp. 1-6.
\bibitem{bvitexture} M. A. Papadopoulos, F. Zhang, D. Agrafiotis and D. Bull, ``A video texture database for perceptual compression and quality assessment," in \textit{IEEE International Conference on Image Processing}, 2015, pp. 2781-2785.
\bibitem{syntex} D. Ma, A. V. Katsenou and D. R. Bull, ``A Synthetic Video Dataset for Video Compression Evaluation," in \textit{IEEE International Conference on Image Processing}, 2019, pp. 1094-1098.
\bibitem{ssim} Z. Wang, A. C. Bovik, H. R. Sheikh and E. P. Simoncelli, ``Image quality assessment: from error visibility to structural similarity," \textit{IEEE Transactions on Image Processing}, vol. 13, no. 4, pp. 600-612, April 2004.
\bibitem{vmaf} Z. Li, A. Aaron, I. Katsavounidis, A. Moorthy, and M. Manohara, ``Toward a practical perceptual video quality metric,” http://techblog. netflix.com/2016/06/toward-practical-perceptual-video.html.
\bibitem{bc4} BlueCrystal Phase 4, https://www.acrc.bris.ac.uk/protected/ bc4-docs/.
\end{thebibliography}
\end{document}